# GRAPHITE CREEP NEGATION DURING FLASH SPARK PLASMA SINTERING UNDER TEMPERATURES CLOSE TO 2000 °C

Charles Manière[a]*, Geuntak Lee[a,b], Joanna McKittrick[b],
Andrey Maximenko[a], and Eugene A. Olevsky[a,c]

(a) Powder Technology Laboratory, San Diego State University, San Diego, USA
(b) Mechanical and Aerospace Engineering, University of California, San Diego, La Jolla, USA
(c) NanoEngineering, University of California, San Diego, La Jolla, USA

**Abstract**

Graphite creep has high importance for applications using high pressures (100 MPa) and temperatures close to 2000 °C. In particular, the new flash spark plasma sintering process (FSPS) is highly sensitive to graphite creep when applied to ultra-high temperature materials such as silicon carbide. In this flash process taking only a few seconds, the graphite tooling reaches temperatures higher than 2000 °C resulting in its irreversible deformation. The graphite tooling creep prevents the flash spark plasma sintering process from progressing further. In this study, a finite element model is used to determine FSPS tooling temperatures. In this context, we explore the graphite creep onset for temperatures above 2000 °C and for high pressures. Knowing the graphite high temperature limit, we modify the FSPS process so that the sintering occurs outside the graphite creep range of temperatures/pressures. 95 % dense silicon carbide compacts are obtained in about 30 s using the optimized FSPS.

---

* Corresponding author: **CM**: Powder Technology Laboratory; College of Engineering; San Diego State University; 5500 Campanile Drive; San Diego, CA 92182-1323
Ph.: +1(619)-594-6329; Fax: +1(619)-594-3599 ; E-mail: charles.maniere@ensicaen.fr

## 1. Introduction

High strength graphite is a material which resists high pressures (~150 MPa), high temperatures (sublimation occurs at 3650 °C) [1] allowing it to serve as a tooling for the pressure assisted sintering of various materials including Ultra-High Temperatures Ceramics (UHTC) [2,3]. UHTC have melting points over 3000 °C, they are resistant to thermal shock, and have high level of refractoriness, chemical inertness, electrical/thermal conductivity and mechanical properties (even at high temperatures). In addition, UHTC materials have strong covalent bonds and low self-diffusion which makes their sintering very difficult [4]. Graphite tools at the maximum allowable pressure and temperature are traditionally employed to sinter these materials *via* processes like Hot Pressing (HP). However, HP cycles are usually long (and conducted under high temperatures) which implies high grain growth and residual intragrain porosity [5,6]. Among all existing approaches, Spark Plasma Sintering (SPS) [7] uses high strength graphite in combination with pulsed electrical current to sinter powdered materials under high pressure and temperature. This process has a very fast heating regime enabling the sintering of nanomaterials [8] (by reducing the grain growth), metals, semiconductors and dielectrics [9–12]. For UHTC, the SPS process represents a very viable technology. The highly responsive heating and shorter sintering cycle of SPS make it suitable to process UHTC materials such as TaC [13], $ZrB_2$ [14], $TiB_2$ [15], etc. Silicon Carbide (SiC) has a melting point below 3000 °C but its sintering is as challenging as of UHTC materials. Pure SiC can be sintered by SPS [16,17]. However, sintering additives such as B-C [18], $Al_3BC_3$ [19], $Al_2O_3$ [20] are traditionally used to help the densification of this material.

In order to sinter the above-mentioned materials without sintering aids, a new SPS approach was developed. High electric current SPS has shown high potential for the densification of SiC (with alumina additive) [21]. At the same time, flash sintering has shown very promising results for the sintering of UHTC and similar high temperature materials [22–24]. Flash sintering is a process which implies an abrupt release of energy in a green specimen and sintering times of few seconds (1-60 s) [25–27]. Since its discovery [28,29], flash sintering has been adapted to different technics like microwave sintering [30] and, in particular, SPS [23,31]. Flash Spark Plasma Sintering (FSPS) has been adapted first in sinter-forging configuration where a pre-sintered specimen was placed in free compression and surrounded by a sacrificial copper bushing (susceptor) to initiate the specimen heating [31]. Using various sacrificial susceptors like copper bushing, graphite felt/foil, FSPS has been employed for consolidation of SiC [31–33], zirconia [34], $ZrB_2$ [24], etc. The sinter-forging configuration is very efficient to eliminate the porosity. However, for controlling the specimen shape, an electrically insulated die can be added to this initial configuration [35]. It was shown that imposing an abrupt electrical current in this die compression flash process, it was possible to flash sinter almost all materials (regardless of their electrical conductivity) from dielectrics to metals [36]. This process has a very stable heating regime due to the combination of the highly concentrated electrical current, the thermal confinement due to the thermal contact resistance [37], and the hybrid heating nature of this sintering approach where the lateral graphite foil dissipates heat.

The latter flash approach is very stable and is studied in the present work to flash spark plasma sinter a pure SiC powder. This method requires very high pressure and temperature (50-100 MPa and 2000-2500 °C) close to the mechanical resistance of graphite. Under these extreme conditions graphite creep occurs impeding the flash spark plasma sintering process. Graphite creep negation is a mandatory step to allow the FSPS of SiC. In this work, we have conducted *in situ* high temperature tests to determine the graphite creep onset temperatures for different pressures. Knowing this information, we studied different approaches to reduce the tooling temperatures while preserving the conditions required to flash spark plasma sinter SiC. The finite element simulation was employed to assist this exploratory work.

## 2. Experiment and method

All the FSPS experiments where carried out using Spark Plasma Sintering system SPSS DR.SINTER Fuji Electronics model 5015. A 45-65 nm SiC nano-size powder (beta, > 99%, US Research Nanomaterials, Inc., Houston, TX) was used for this study. The graphite punches, spacers and die were made from EDM4 graphite (Poco Graphite, Inc., Texas, USA) which among all grades has the best mechanical performances. The die surface temperature was measured by a pyrometer (Chino, IR-AHS2). We also measured the punch temperature by a sacrificial K type thermocouple which indicated the temperature profile of the punch at the onset of the flash phenomena before its destruction at ~1400 °C. The FSPS configuration is similar to our previous work [36] and uses 10 mm punches and a BN electrically insulated graphite dies which concentrate the electrical current flux in the powder and the surrounding graphite foil. A constant 90 MPa pressure was applied. FSPS is imposed by an electrical current profile where the specimen is preheated essentially by the graphite foil to raise its electrical conductivity; after that, an abrupt current peak is achieved up to the stabilization of the punch displacement indicating the end of the sintering. The well- grinded and polished samples were etched for 23 min with the Murakami's reagent (Solution of 200mL DI water, 10g KOH and 10g $K_3Fe(CN)_6$) which was located in the glass beaker in the boiling water. The etched specimens were analyzed by the scanning electron microscopy (FEI Quanta 450, USA).

In this study, the FSPS of SiC is first analyzed in vacuum and argon atmosphere. This shows the sintering response of SiC and the impact of cooling fluxes by convection/radiation on the graphite tooling creep. The improvement of the FSPS of the SiC specimen was tested using a high temperature forging experiment.

The real temperature field during the FSPS is difficult to obtain experimentally. A finite element simulation is investigated to estimate the FSPS temperatures. This simulation takes into account the Joule heating and the cooling fluxes by surface to surface thermal radiation and by convection (for FSPS experiment in argon atmosphere). The problem formulation and boundary conditions of the Joule heating and thermal radiation fluxes can be found in our previous work [37–39]. The formulation of the natural convection during the FSPS experiment in argon atmosphere uses Navier-Stokes equations, and the formulation of the convective problem can be found in our previous work [40]. After the estimation of the tooling temperatures, the onset of the graphite creep is determined using 10 mm punches *via* a pyrometer measurement at the maximum temperature point and for different applied pressures (from 50 to 100 MPa). For these tests, a 100 K/min ramp is forced up to the detection of graphite creep in the displacement curves under constant pressure. When the graphite creep onset temperatures and pressure are known, the alternative FSPS routines of SiC were investigated: one with two pressure steps and another with smaller punches, these experiments aimed to avoid the graphite tooling creep in order to achieve a full densification of SiC.

## 3. Results and discussions

### *3.1. Flash spark plasma sintering under vacuum and argon atmospheres*

The results of the FSPS of SiC in argon and vacuum atmospheres are reported in figure 1. Figure 1 shows the abrupt electric current profile imposed on the specimen and the temperature/displacement response. The electric current peak generates an abrupt thermal response in the punches and the die with heating rates close to 4000 K/min in the punch and 300 K/min in the die. Compared to the die (electrically insulated), the punch has a high heating reactivity [41] because of the high electrical current concentration. Sintering times of 20 s and 35 s were obtained for "argon" and "vacuum" experiments, respectively. The main differences in the temperatures curves are noticeable for the punches, where the onset of the abrupt temperatures profile appears at 900 °C in argon and at 1200 °C in vacuum. The obtained results seem to indicate higher cooling fluxes at the punch surface for the FSPS experiments in argon atmosphere. This may be explained by the convection motion in the cavity which is not present in vacuum. The finite element simulation helps understand this phenomenon. Finally, the displacement curves seem to indicate a different sintering profile for argon atmosphere and vacuum tests. However, the final sintered relative densities of the obtained pellets are close to 88 % and are not improved by holding at high temperatures. The difference in displacement curves is explained by the graphite creep occurring under high temperatures. Figure 2 indicates that the graphite punches experience a significant amount of creep during the flash process. The diameter of the punches increases up to the point when they discontinue the sliding displacement relative the die which stops the sintering of SiC at about 88 % of relative density.

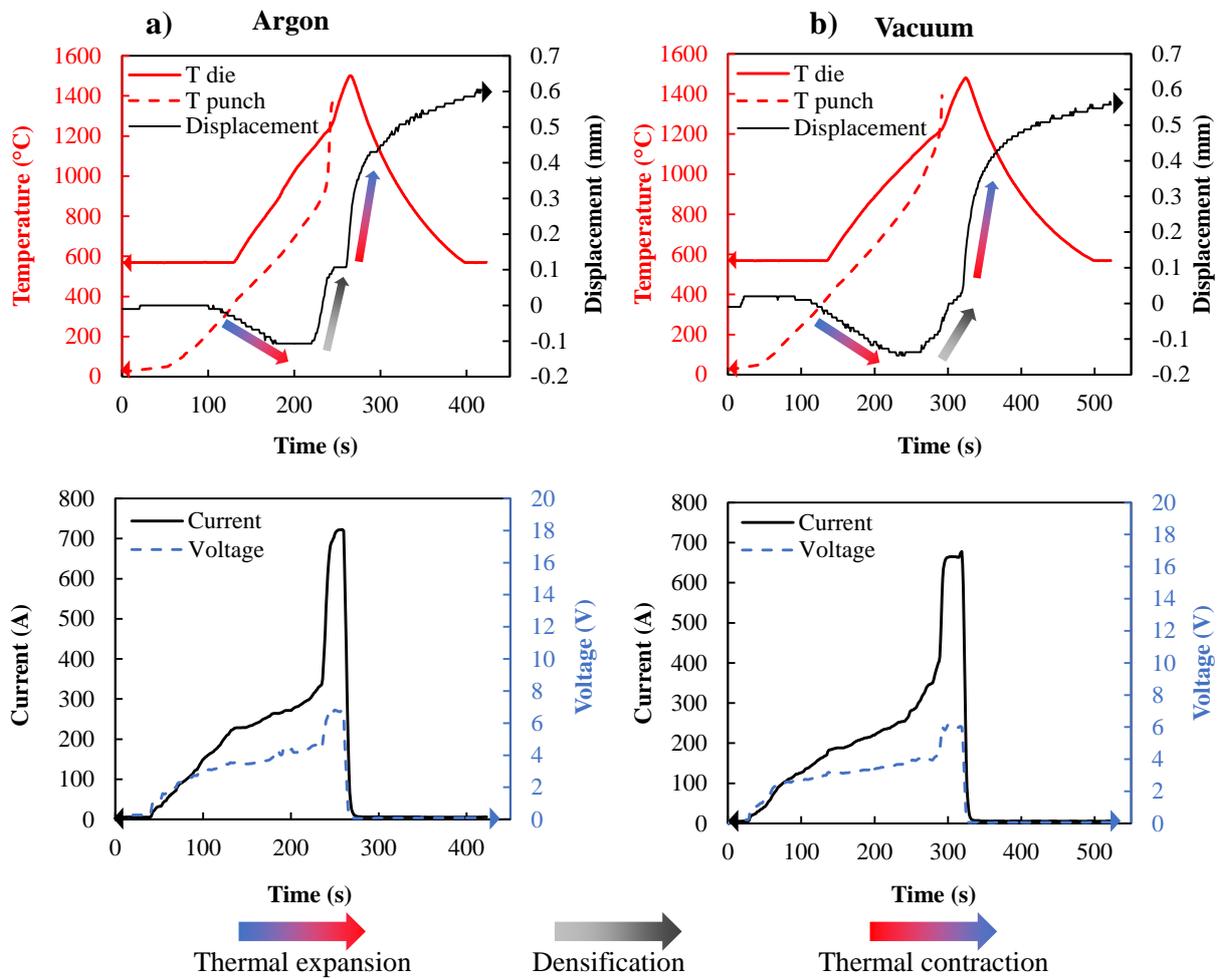

*Figure 1 Temperatures, displacement, electrical current and voltage experimental curves for the flash spark plasma sintering experiments in a) argon and b) vacuum.*

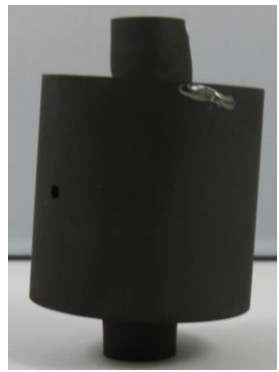

*Figure 2 Photograph of the deformed punches after the flash spark plasma sintering tests.*

*3.2. Finite element simulation of the flash spark plasma sintering process*

In the previous section, a significant creep deformation is observed in the punches (figure 2). It is difficult to estimate the temperature of the punches as the K-type thermocouple located in the creep zone is quickly destroyed by the high temperatures of FSPS. The finite element simulation of both FSPS processes has been investigated to estimate the temperatures in the tooling and the specimen.

Figure 3 shows the simulated temperature curves (and fields) for the FSPS experiments in argon and in vacuum. The temperature curves in figure 3a show the abrupt profile of FSPS with a high heating response in the SiC specimen, the punches and a delayed heating response in the die. The thermal contact resistance [37] and the absence of electrical currents in the die explains this heating delay. The curves for argon and vacuum experiments are close even though the punch ending temperature in argon is decreased from 2400 °C to 2000 °C due to the presence of cooling fluxes by convection in the cavity. The convective motion implies high gas velocities (0.4 m/s in figure 3b) which together with thermal radiation represent an important source of the heat removal in the tooling (as high as during microwave heating [42]).

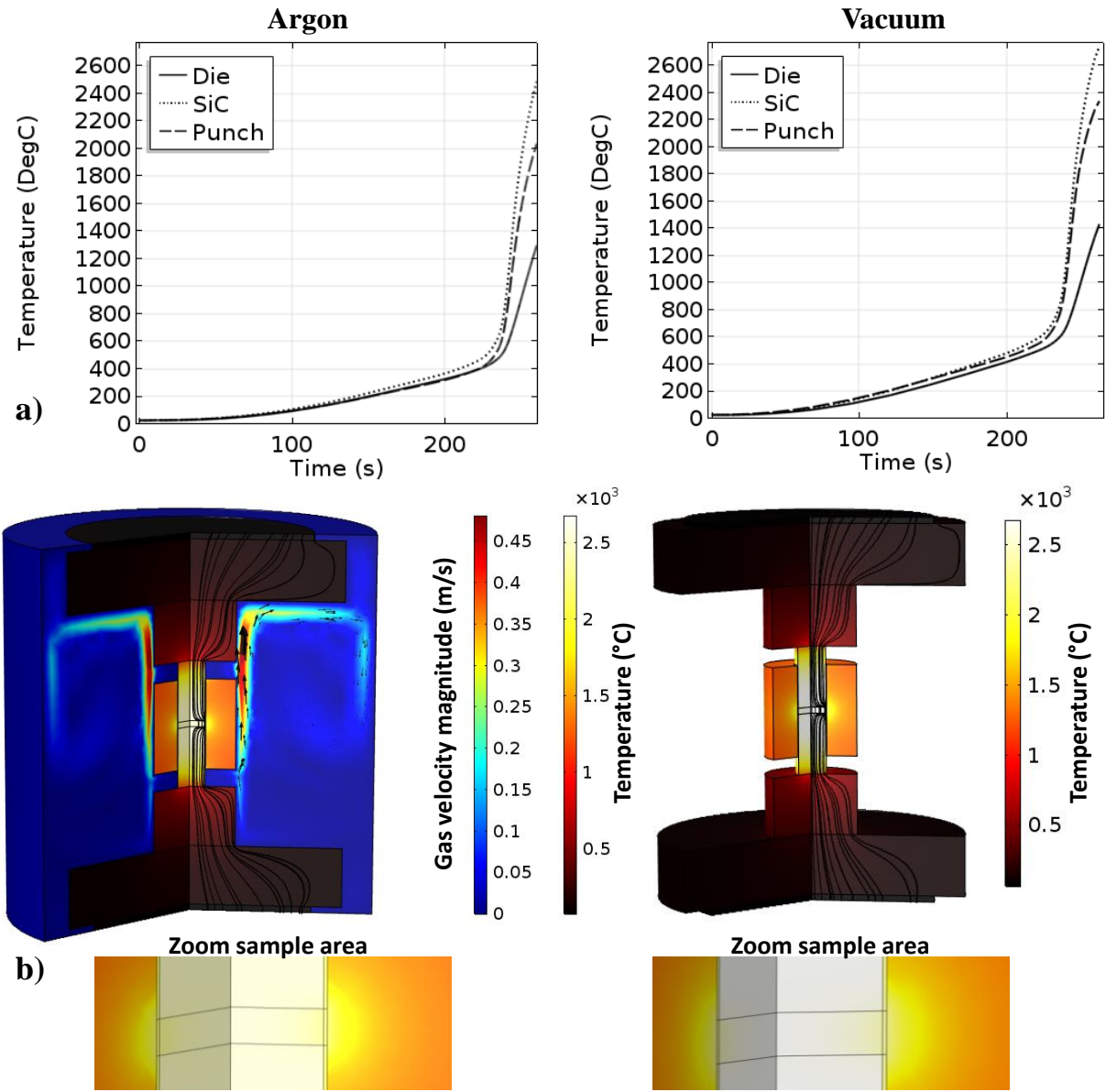

*Figure 3 Simulated flash spark plasma sintering tests in argon and vacuum, a) average punch, die and specimen temperature curves, b) temperature field and electric current lines at the maximum sintering temperatures.*

As shown in the simulated temperatures field in figure 3b, the SiC specimen temperatures are higher in the center due to a cooling of the die at the specimen edge. The simulations predict that, during few seconds, SiC temperatures are as high as 2600 °C, a temperature level high enough for the densification of the powder under pressure. However, the punches outside the die have temperatures around 2000 °C, which in combination with the pressure (90 MPa) implies creep.

*3.3. High temperature forging of flash sintered specimens*

In this section, we describe our attempt to fully densify the 88 % FSPS SiC specimen by adding a high temperature forging step. The 10 mm specimen was placed in a 15 mm inner diameter die and we manually increased the heating up to achieving the temperatures close to 2000 °C and the pressures up to the detection of the creep displacement. Figure 4 shows temperatures, displacement, electrical current, pressure curves, the configuration scheme and the photographs of the heating and cooling specimens.

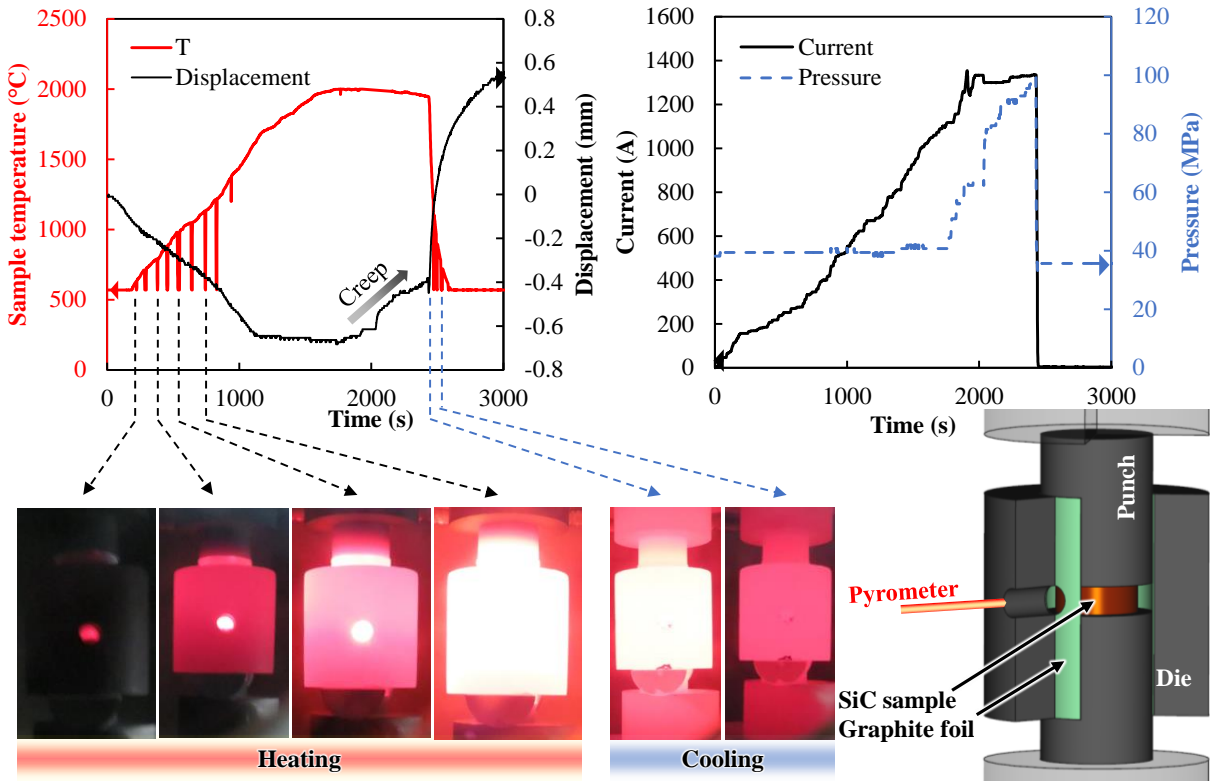

*Figure 4 High temperature forging experiment on the flash spark plasma sintered specimens. The photograph of the forging configuration during the heating indicates that the specimen has an electrical conductivity allowing a preferential heating in the sample rather than in the*

graphite die (not electrically insulated in this configuration). The breakdown in the temperature curves (in red) corresponds to the photograph taken during the heating cycle in front of the pyrometer (which temporarily stops the temperatures measurement). After the detection of creep, the pressure was increased to a value close to 100 MPa and the samples were cooled. Despite the high temperature and pressure, the specimen densification was not improved and the punches were deformed. The densification of SiC at high temperature is therefore very challenging and requires avoiding the conditions favoring the graphite tooling creep.

*3.4. Determination of high temperature graphite creep onset*

There are no well-defined data on the onset temperature/pressure conditions for high strength graphite creep at high temperatures. This information is essential to determine optimal sintering conditions for SiC or UHTC materials which do not implicate any damage of the graphite tooling. To investigate the temperature/pressure limit of EDM4 graphite, we imposed a 100 K/min heating on a 10 mm diameter graphite punch up to the creep detection and for the constant pressures of 50, 75 and 100 MPa. The results are reported in figure 5. The deformed punches are shown in the lower photograph. The barrel shape indicates that the maximum temperature is located in the center of the specimen where the pyrometer measurement is made. The onset creep data correspond then to the punches' maximum temperatures. The displacement curves indicate a linear decrease during the heating ramp corresponding to the thermal expansion of graphite. The creep starts when a rupture is observed in this linear decrease. The displacement curve at 100 MPa indicates that there are two creep onsets: one corresponds to a small creep displacement at 1600 °C and another one corresponds to a severe creep at 2050 °C. For 75 MPa the first creep onset is at 1900 °C and the second is at 2200 °C. For 50 MPa, it is 2300 °C and 2500 °C, respectively. These data can be used to avoid the FSPS graphite tooling creep. Since creep causes irreversible deformation, even a week creep temperatures/pressures range should be avoided. Indeed, the week creep of the FSPS tooling may allow sintering because of small tooling deformations but repetitive experiments would quickly damage the graphite tools and deteriorate the FSPS experimental reproducibility. Therefore all creep areas in figure 5 are considered to be a "no-go zone".

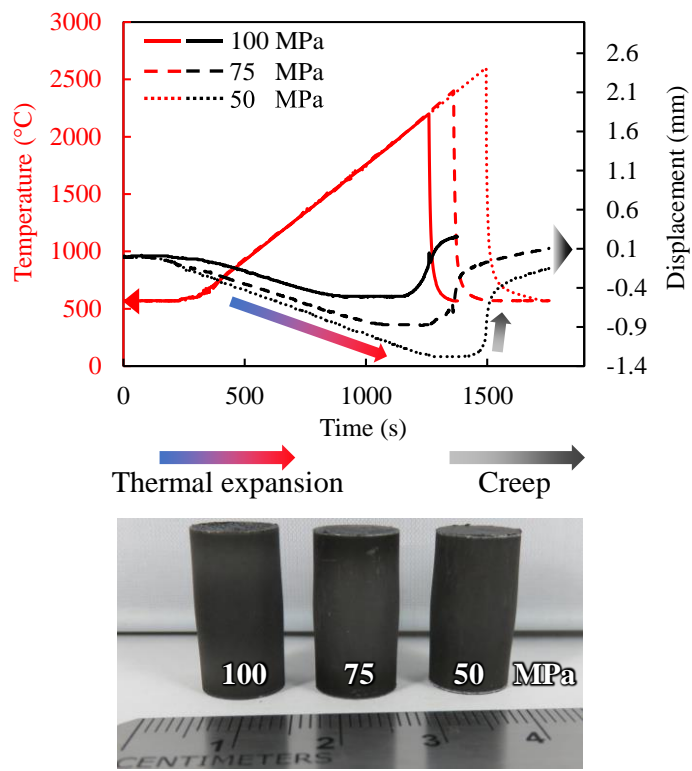

*Figure 5 Graphite creep onset determination for 10 mm diameter punches.*

*3.5. Improved flash spark plasma sintering experiments*

In order to carry out the FSPS of SiC without graphite tooling creep, the new tests locate the pyrometer temperature measurement on the punches (in the area where graphite creep was detected, see figure 2). In this way, it is possible to reduce the pressure or stop the heating when the measured temperature and the corresponding pressure reach the "no-go zone". Two strategies were used to avoid the "no-go zone" of graphite creep. One decreases the pressure when the temperature approaches the onset creep temperature ("2 step pressure"). The other consists of using smaller punches to decrease the punches' temperatures by increasing the cooling of the spacer ("small punches"). Indeed, the profile of the graphite creep in figure 2 shows that the area of the punches in contact with the spacer is not deformed because they are cooled by the spacer. The "small punches" approach intends to extend this area thereby reducing the punch temperatures and the creep phenomenon. The results are reported in figure 6.

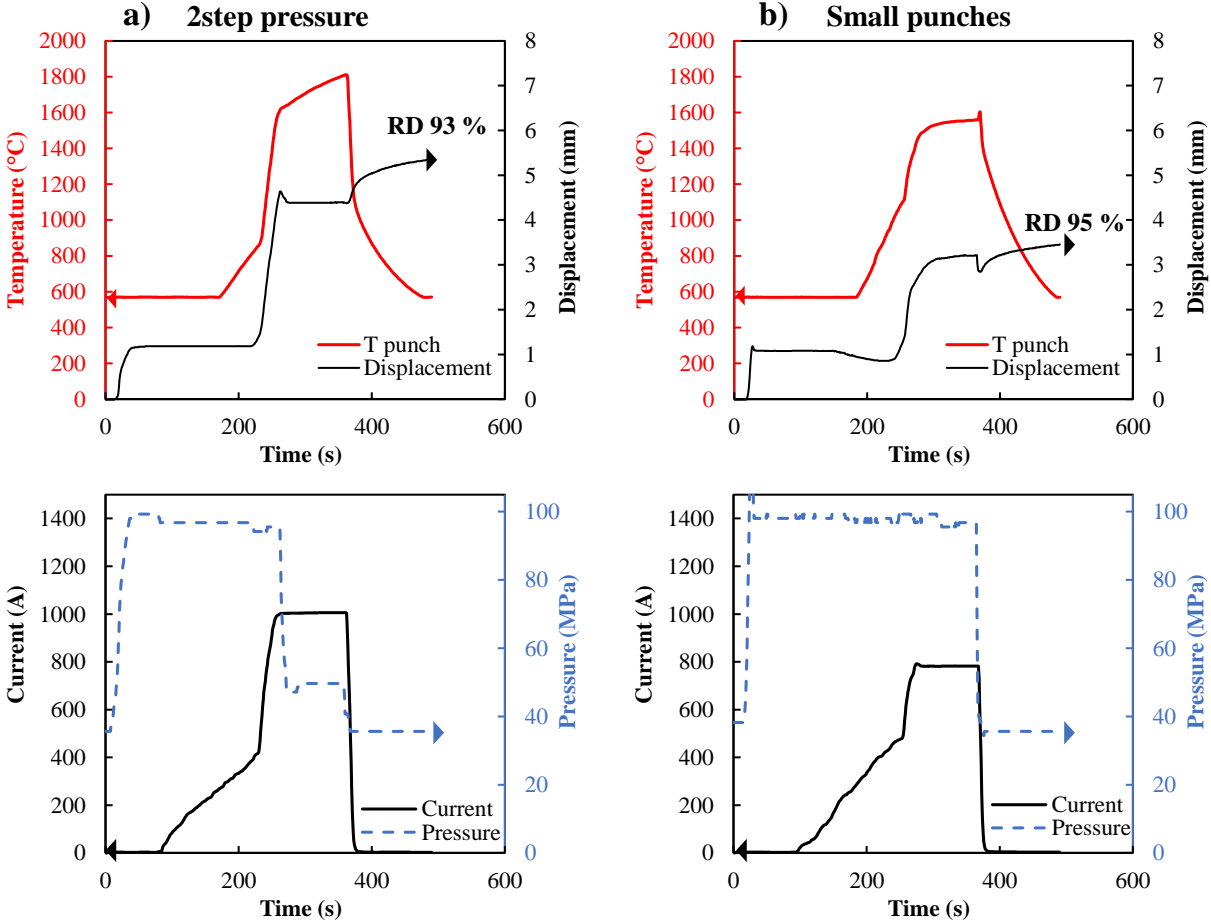

*Figure 6 Optimized flash spark plasma sintering experiments using a) two pressure steps and b) smaller punches.*

For the "2-step pressure" approach, one can see that the initial pressure of 100 MPa was decreased to 50 MPa when the punch temperature started approaching the creep onset temperature of 1600 °C (see figure 5). For both approaches, the sintering displacement amplitude is larger than in the previous tests (figure 1) and value of SiC densification exceeds 90 % (93 % for "2-step pressure" and 95 % for "small punches"). The microstructures obtained in the center and the edge of the sample are reported in figure 7. For the "2 step pressure" FSPS approach the microstructure is relatively uniform in the pellet even if larger grains are observed in the center indicating a higher temperature in this area. For the "small punches" FSPS configuration, even higher temperatures seem to be present in the center of the specimen. Larger grains are observed and in the upper central area; the hot spot seems to have generated larger grains with an inter-grain liquid silicon rich phase (Si 69.65 % and C 30.35 % by EDX). Similar phenomena of SiC dissociation are reported for high temperatures (2550 °C) [43]. This confirms the possible presence of a hot spot common for the Negative Temperature Coefficient resistivity (NTC) materials where higher temperatures amplify the local dissipation and the electric current concentration in these areas [44]. The overall specimen has a higher densification than after using the "2-Step pressure" approach mainly due to a well-densified area at the edge of the specimen.

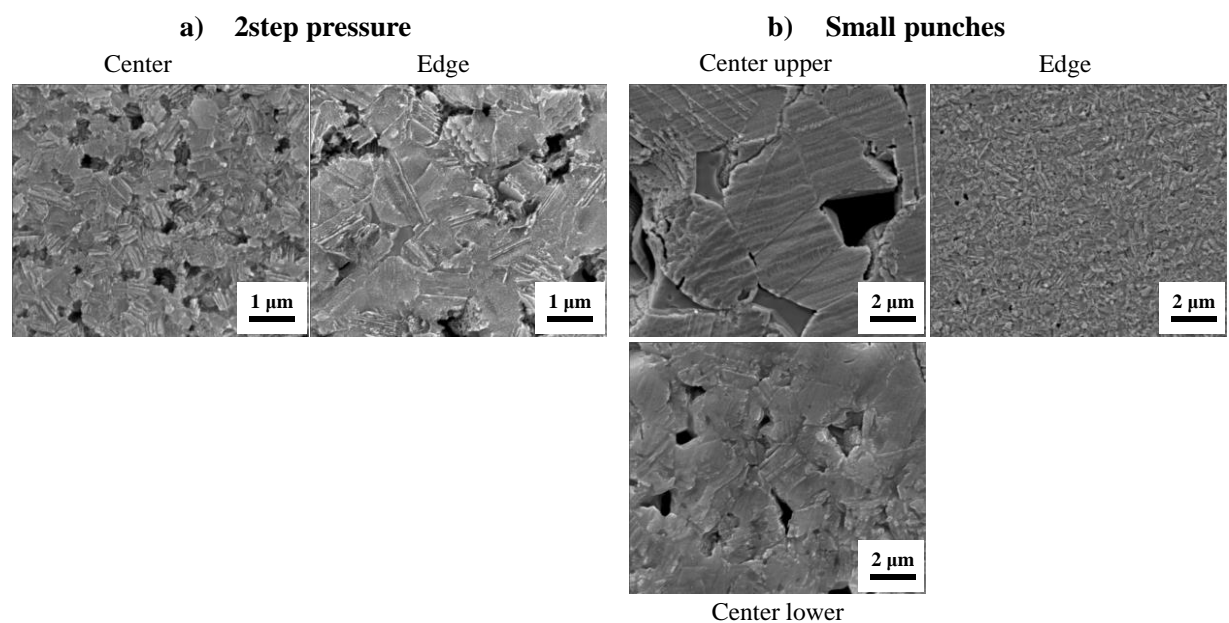

*Figure 7 Polished and etched microstructures obtained for the flash spark plasma sintered specimens in a) "2-step pressure" and b) "small punches" configurations.*

## 4. Conclusion

Flash spark plasma sintering of silicon carbide is a very challenging processing technique requiring very high temperatures (~2000 °C) and pressures. Furthermore, the abrupt nature of this "flash" process favors the appearance of significant peaks of temperatures which can activate the creep of the graphite tools and prevent the full densification of silicon carbide (even after additional forging treatment). A finite element simulation reveals that the tooling temperatures can easily exceed 2000 °C. Identifying the temperature and pressure limits of high strength graphite is a key issue for this SiC flash processing. To achieve this objective, we have determined the FSPS graphite tooling creep onset for different pressures and temperatures. These data were used to optimize the flash spark plasma sintering by introducing a new processing method which approaches the graphite resistance limit allowing the FSPS of SiC that can achieve 95 % relative density during processing time of about 30-40 s.

The developed process still needs the improvement of the homogeneity of the obtained microstructures which seem to indicate higher temperatures in the center of the specimens. Higher grain sizes and the presence of silicon rich inter-grain phase was observed in the small punch experiment. This suggests the presence of a high local temperature (hot spot) explaining the graphite dissociation phenomena that should be mitigated to make this process suitable for practical applications.

**Acknowledgements**

The support of the NSF Division of Materials Research (Grant DMR-1900876) is gratefully acknowledged.

# References

[1]   R.K. Carlson, J.J. Ferritto, Manufacture of high density, high strength isotropic graphite, US4226900A, 1978. https://patents.google.com/patent/US4226900.

[2]   J.F. Justin, A. Jankowiak, Ultra High Temperature Ceramics : Densification,Properties and Thermal Stability, AerospaceLab. (2011) 1–11. doi:hal-01183657.

[3]   R.M. German, Sintering Theory and Practice, Wiley, Wiley, 1996. http://www.wiley.com/WileyCDA/WileyTitle/productCd-047105786X.html.

[4]   W.G. Fahrenholtz, G.E. Hilmas, I.G. Talmy, J.A. Zaykoski, Refractory Diborides of Zirconium and Hafnium, J. Am. Ceram. Soc. 90 (2007) 1347–1364. doi:10.1111/j.1551-2916.2007.01583.x.

[5]   S.-Q. Guo, Densification of ZrB2-based composites and their mechanical and physical properties: A review, J. Eur. Ceram. Soc. 29 (2009) 995–1011. doi:10.1016/j.jeurceramsoc.2008.11.008.

[6]   O. Guillon, J. Gonzalez-Julian, B. Dargatz, T. Kessel, G. Schierning, J. Räthel, M. Herrmann, Field-Assisted Sintering Technology/Spark Plasma Sintering: Mechanisms, Materials, and Technology Developments, Adv. Eng. Mater. 16 (2014) 830–849. doi:10.1002/adem.201300409.

[7]   E.A. Olevsky, D. V. Dudina, Field-Assisted Sintering, Springer International Publishing, Cham, 2018. doi:10.1007/978-3-319-76032-2.

[8]   F. Maglia, I.G. Tredici, U. Anselmi-Tamburini, Densification and properties of bulk nanocrystalline functional ceramics with grain size below 50nm, J. Eur. Ceram. Soc. 33 (2013) 1045–1066. doi:10.1016/j.jeurceramsoc.2012.12.004.

[9]   R. Orrù, R. Licheri, A.M. Locci, A. Cincotti, G. Cao, Consolidation/synthesis of materials by electric current activated/assisted sintering, Mater. Sci. Eng. R Reports. 63 (2009) 127–287. doi:10.1016/j.mser.2008.09.003.

[10]  W. Chen, U. Anselmi-Tamburini, J.E. Garay, J.R. Groza, Z.A. Munir, Fundamental investigations on the spark plasma sintering/synthesis process, Mater. Sci. Eng. A. 394 (2005) 132–138. doi:10.1016/j.msea.2004.11.020.

[11]  R.K. Bordia, S.-J.L. Kang, E.A. Olevsky, Current understanding and future research

directions at the onset of the next century of sintering science and technology, J. Am. Ceram. Soc. 100 (2017) 2314–2352. doi:10.1111/jace.14919.

[12] S.H. Risbud, Y.-H. Han, Preface and historical perspective on spark plasma sintering, Scr. Mater. 69 (2013) 105–106. doi:10.1016/j.scriptamat.2013.02.024.

[13] X. Zhang, G.E. Hilmas, W.G. Fahrenholtz, Densification and mechanical properties of TaC-based ceramics, Mater. Sci. Eng. A. 501 (2009) 37–43. doi:10.1016/j.msea.2008.09.024.

[14] E. Wuchina, E. Opila, W. Fahrenholtz, I. Talmy, UHTCs: Ultra-High Temperature Ceramic Materials for Extreme Environment Applications, Electrochem. Soc. Interface. (2007) 30–36.

[15] Z.H. Zhang, X.B. Shen, F.C. Wang, S.K. Lee, L. Wang, Densification behavior and mechanical properties of the spark plasma sintered monolithic TiB2 ceramics, Mater. Sci. Eng. A. 527 (2010) 5947–5951. doi:10.1016/j.msea.2010.05.086.

[16] A. Reau, F. Tenegal, J. Galy, Process for preparing a silicon carbide part without the need for any sintering additives, US8871141B2, 2008. https://patents.google.com/patent/US8871141.

[17] A. Lara, A.L. Ortiz, A. Muñoz, A. Domínguez-Rodríguez, Densification of additive-free polycrystalline β-SiC by spark-plasma sintering, Ceram. Int. 38 (2012) 45–53. doi:10.1016/j.ceramint.2011.06.036.

[18] T.A. Yamamoto, T. Kondou, Y. Kodera, T. Ishii, M. Ohyanagi, Z.A. Munir, Mechanical Properties of β-SiC Fabricated by Spark Plasma Sintering, J. Mater. Eng. Perform. 14 (2005) 460–466. doi:10.1361/105994905X56250.

[19] S.-H. Lee, H. Tanaka, Y. Kagawa, Spark plasma sintering and pressureless sintering of SiC using aluminum borocarbide additives, J. Eur. Ceram. Soc. 29 (2009) 2087–2095. doi:10.1016/j.jeurceramsoc.2008.12.006.

[20] M.D. Unlu, G. Goller, O. Yucel, F.C. Sahin, The Spark Plasma Sintering of Silicon Carbide Ceramics Using Alumina, Acta Phys. Pol. A. 125 (2014) 257–259. doi:10.12693/APhysPolA.125.257.

[21] B. Román-Manso, M. Belmonte, M.I. Osendi, P. Miranzo, Effects of Current Confinement on the Spark Plasma Sintering of Silicon Carbide Ceramics, J. Am. Ceram. Soc. 98 (2015) 2745–2753. doi:10.1111/jace.13678.

[22] E. Zapata-Solvas, S. Bonilla, P.R. Wilshaw, R.I. Todd, Preliminary investigation of flash sintering of SiC, J. Eur. Ceram. Soc. 33 (2013) 2811–2816. doi:10.1016/j.jeurceramsoc.2013.04.023.

[23] G. Cabouro, S. Le Gallet, S. Chevalier, E. Gaffet, Y. Grin, F. Bernard, Dense Mosi2 produced by reactive flash sintering: Control of Mo/Si agglomerates prepared by high-energy ball milling, Powder Technol. 208 (2011) 526–531. doi:10.1016/j.powtec.2010.08.054.

[24] S. Grasso, T. Saunders, H. Porwal, O. Cedillos-Barraza, D.D. Jayaseelan, W.E. Lee, M.J. Reece, Flash Spark Plasma Sintering (FSPS) of Pure ZrB 2, J. Am. Ceram. Soc. 97 (2014) 2405–2408. doi:10.1111/jace.13109.

[25] M. Yu, S. Grasso, R. Mckinnon, T. Saunders, M.J. Reece, Review of flash sintering: materials, mechanisms and modelling, Adv. Appl. Ceram. 116 (2017) 24–60. doi:10.1080/17436753.2016.1251051.

[26] C.E.J. Dancer, Flash sintering of ceramic materials, Mater. Res. Express. 3 (2016) 102001. doi:10.1088/2053-1591/3/10/102001.

[27] M. Biesuz, V.M. Sglavo, Flash sintering of ceramics, J. Eur. Ceram. Soc. 39 (2019) 115–143. doi:10.1016/j.jeurceramsoc.2018.08.048.

[28] R. Raj, M. Cologna, A.L.G. Prette, V. Sglavo, Methods of flash sintering, Patent US 20130085055 A1, US 20130085055 A1, 2013. https://www.google.com/patents/US20130085055.

[29] M. Cologna, B. Rashkova, R. Raj, Flash Sintering of Nanograin Zirconia in <5 s at 850°C, J. Am. Ceram. Soc. 93 (2010) 3556–3559. doi:10.1111/j.1551-2916.2010.04089.x.

[30] Y. Bykov, S. Egorov, A. Eremeev, V. Kholoptsev, I. Plotnikov, K. Rybakov, A. Sorokin, On the Mechanism of Microwave Flash Sintering of Ceramics, Materials (Basel). 9 (2016) 684. doi:10.3390/ma9080684.

[31] E.A. Olevsky, S. Rolfing, Y.S. Lin, A. Maximenko, Flash spark-plasma sintering of SiC powder, in: 10th Pacific Rim Conf. Ceram. Glas. Technol. Coronado, CA, Coronado, 2013: p. 32.

[32] S. Grasso, T. Saunders, H. Porwal, B. Milsom, A. Tudball, M. Reece, Flash Spark Plasma Sintering (FSPS) of α and β SiC, J. Am. Ceram. Soc. 99 (2016) 1534–1543.

doi:10.1111/jace.14158.

[33] E.A. Olevsky, S.M. Rolfing, A.L. Maximenko, Flash (Ultra-Rapid) Spark-Plasma Sintering of Silicon Carbide, Sci. Rep. 6 (2016) 33408. doi:10.1038/srep33408.

[34] O. Vasylkiv, H. Borodianska, Y. Sakka, D. Demirskyi, Flash spark plasma sintering of ultrafine yttria-stabilized zirconia ceramics, Scr. Mater. 121 (2016) 32–36. doi:10.1016/j.scriptamat.2016.04.031.

[35] E. Zapata-Solvas, D. Gómez-García, A. Domínguez-Rodríguez, R.I. Todd, Ultra-fast and energy-efficient sintering of ceramics by electric current concentration, Sci. Rep. 5 (2015) 8513. doi:10.1038/srep08513.

[36] C. Manière, G. Lee, E.A. Olevsky, All-Materials-Inclusive Flash Spark Plasma Sintering, Sci. Rep. 7 (2017) 15071. doi:10.1038/s41598-017-15365-x.

[37] C. Manière, L. Durand, E. Brisson, H. Desplats, P. Carré, P. Rogeon, C. Estournès, Contact resistances in spark plasma sintering: From in-situ and ex-situ determinations to an extended model for the scale up of the process, J. Eur. Ceram. Soc. 37 (2017) 1593–1605. doi:10.1016/j.jeurceramsoc.2016.12.010.

[38] C. Manière, E. Torresani, E. Olevsky, Simultaneous Spark Plasma Sintering of Multiple Complex Shapes, Materials (Basel). 12 (2019) 557. doi:10.3390/ma12040557.

[39] C. Manière, T. Zahrah, E.A. Olevsky, Fully coupled electromagnetic-thermal-mechanical comparative simulation of direct vs hybrid microwave sintering of 3Y-ZrO2, J. Am. Ceram. Soc. 100 (2017) 2439–2450. doi:10.1111/jace.14762.

[40] C. Manière, T. Zahrah, E.A. Olevsky, Fluid dynamics thermo-mechanical simulation of sintering: Uniformity of temperature and density distributions, Appl. Therm. Eng. 123 (2017) 603–613. doi:10.1016/j.applthermaleng.2017.05.116.

[41] C. Manière, G. Lee, E.A. Olevsky, Proportional integral derivative, modeling and ways of stabilization for the spark plasma sintering process, Results Phys. 7 (2017) 1494–1497. doi:10.1016/j.rinp.2017.04.020.

[42] S. V. Egorov, K.I. Rybakov, V.E. Semenov, Y. V. Bykov, O.N. Kanygina, E.B. Kulumbaev, V.M. Lelevkin, Role of convective heat removal and electromagnetic field structure in the microwave heating of materials, J. Mater. Sci. 42 (2007) 2097–2104. doi:10.1007/s10853-006-0157-x.

[43] H.K. Henisch, R. Roy, Silicon Carbide–1968, Elsevier, 1969. doi:10.1016/C2013-0-01599-X.

[44] C. Manière, T. Zahrah, E.A. Olevsky, Inherent heating instability of direct microwave sintering process: Sample analysis for porous 3Y-ZrO2, Scr. Mater. 128 (2017) 49–52. doi:10.1016/j.scriptamat.2016.10.008.

**Figure captions**

Figure 1 Temperatures, displacement, electrical current and voltage experimental curves for the flash spark plasma sintering experiments in a) argon and b) vacuum.

Figure 2 Photograph of the deformed punches after the flash spark plasma sintering tests.

Figure 3 Simulated flash spark plasma sintering tests in argon and vacuum, a) average punch, die and specimen temperature curves, b) temperature field and electric current lines at the maximum sintering temperatures.

Figure 4 High temperature forging experiment on the flash spark plasma sintered specimens.

Figure 5 Graphite creep onset determination for 10 mm diameter punches.

Figure 6 Optimized flash spark plasma sintering experiments using a) two pressure steps and b) smaller punches.

Figure 7 Polished and etched microstructures obtained for the flash spark plasma sintered specimens in a) "2-step pressure" and b) "small punches" configurations.